\documentclass[preprint,amsmath,amssymb,aps,superscriptaddress]{revtex4-1}

\usepackage{graphicx}
\usepackage{dcolumn}
\usepackage{bm}
\usepackage{color}
\usepackage{multirow}
\usepackage{epstopdf}

\begin{document}

\preprint{APS/123-QED}

\title{Role of crystal structure and junction morphology on interface thermal conductance}

\author{Carlos A. Polanco}
\email{cap3fe@virginia.edu}
\affiliation{Department of Electrical and Computer Engineering, University of Virginia, Charlottesville,VA-22904.}%

\author{Rouzbeh Rastgarkafshgarkolaei}
\affiliation{Department of Mechanical and Aerospace Engineering, University of Virginia, Charlottesville,VA-22904.}%

\author{Jingjie Zhang}
\affiliation{Department of Electrical and Computer Engineering, University of Virginia, Charlottesville,VA-22904.}%

\author{Nam Q. Le}
\affiliation{Department of Mechanical and Aerospace Engineering, University of Virginia, Charlottesville,VA-22904.}%

\author{Pamela M. Norris}
\affiliation{Department of Mechanical and Aerospace Engineering, University of Virginia, Charlottesville,VA-22904.}%

\author{Patrick E. Hopkins}
\affiliation{Department of Mechanical and Aerospace Engineering, University of Virginia, Charlottesville,VA-22904.}%

\author{Avik W. Ghosh}
\affiliation{Department of Electrical and Computer Engineering, University of Virginia, Charlottesville,VA-22904.}%

\date{\today}

\begin{abstract}
We argue that the relative thermal conductance between interfaces with different morphologies is controlled by crystal structure through $M_{min}/M_c > 1$, the ratio between the {\it minimum mode} count on either side $M_{min}$, and the {\it conserving modes} $M_c$ that preserve phonon momentum transverse to the interface. Junctions with an added homogenous layer, ``uniform'', and ``abrupt'' junctions are limited to $M_c$ while junctions with interfacial disorder, ``mixed'', exploit the expansion of mode spectrum to $M_{min}$. In our studies with cubic crystals, the largest enhancement of conductance from ``abrupt'' to ``mixed'' interfaces seems to be correlated with the emergence of voids in the conserving modes, where $M_c = 0$. Such voids typically arise when the interlayer coupling is weakly dispersive, making the bands shift rigidly with momentum. Interfacial mixing also increases alloy scattering, which reduces conductance in opposition with the mode spectrum expansion. Thus the conductance across a ``mixed'' junction does not always increase relative to that at a ``uniform'' interface. 
\end{abstract}

\pacs{Valid PACS appear here}
\maketitle


\section{Introduction}

For over half a century, the thermal energy flow across solid-solid interfaces has been studied with only partial understanding of the underlying processes \cite{Cahill2014, Cahill2003, Kim2007}. A microscopic understanding of these interfacial thermal processes requires deconstructing thermal  interfacial conductance, which brings many challenges, including consideration of a broad spectrum of interacting dispersive phonons, varying mean free paths, and additional phonon interactions with defects, impurities and other interfacial imperfections \cite{Hopkins2013a}. Moreover, as the spacing between two interfaces reduces to distances on the order of the phonon coherence length, wave interference and coherent transport contribute to the thermal resistance in a non-additive fashion \cite{Simkin2000,Ravichandran2014,Latour2014,Wang2014}. 

\begin{figure}[t]
	\centering
	\includegraphics[width=86mm]{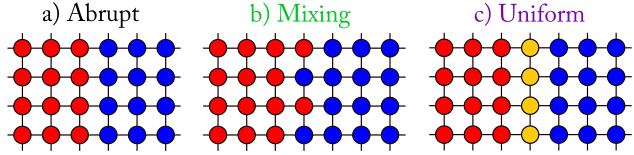}
	\caption{a) Abrupt interface. b) Interface with random atomic mixing (mixed interface). c) Interface with an added homogeneous layer at the junction (uniform interface).}
	\label{fig_cases}
\end{figure}

Early models of interfacial thermal conductance focused on the effect of acoustic matching \cite{Little1959}, nonlinear dispersion \cite{Young1989}, and bonding \cite{Young1989} on perfectly abrupt interfaces (Fig.~\ref{fig_cases}a). Interfacial imperfections were later included in the diffuse mismatch model (DMM) as sources of diffuse scattering \cite{Swartz1989}. Although this model is widely used, it does not account for atomistic interfacial details \cite{Beechem2007}, which have been shown to affect interface conductance measurements \cite{Hopkins2013a}.

One interfacial imperfection is random atomic mixing (Fig.~\ref{fig_cases}b), which can be a frequent byproduct of nanostructure fabrication. The addition of random atoms at an abrupt interface generates two effects in the harmonic regime: 1) it changes phonon transmission \cite{Kozorezov1998, Fagas1999, Zhao2009, Zhao2005}; and, 2) it couples phonons with different transverse wavevectors ($k_\perp$) by breaking the translational symmetry at the interfacial plane \cite{Kechrakos1990}. Some papers focused on the effect of mixing on transmission (Effect 1) and showed the importance of the frequency dependence \cite{Kozorezov1998}, the correlation length of the random distribution \cite{Fagas1999,Zhao2009} and the acoustic-optic coupling \cite{Zhao2005}.

Interestingly, the papers that focused on the coupling of phonons (Effect 2) noted that the thermal conductance of the mixed interface was larger than that of the abrupt interface for a simple cubic crystal interface \cite{Kechrakos1990,Kechrakos1991} and for a Si/heavy-Si interface \cite{Tian2012}. Moreover, Kechrakos \cite{Kechrakos1991} noticed that the conductance of the mixed interface was even larger than the conductance of an interface with an added homogeneous atomic layer (Fig.~\ref{fig_cases}c), which we call a uniform interface. Those results suggest that adding disorder at interfaces increases conductance, contrary to bulk materials where adding disorder has the opposite effect. However, the role of crystal structure on the conductance of the abrupt, mixed and uniform interfaces remains unclear.

In this paper, we demonstrate that the relative conductance between interfaces with different morphologies (Fig.~\ref{fig_cases}) depends on crystal structure. For the systems considered in this study, adding a unit cell monolayer of mixing to an abrupt interface always enhances the interfacial conductance, but the extent varies over an order of magnitude according to the crystal structure (Fig.~\ref{fig_matvsmix}). Furthermore, the conductance across a mixed interface does not always increase relative to that at a uniform interface. In fact, while the conductance increases for simple cubic (SC) and diamond cubic (DC) crystal structures, it decreases for face centered cubic (FCC) crystals. This suggests that the commonly invoked virtual crystal approximation, which models the mixed interface as a uniform interface, alternatively overestimates or underestimates the effect of interfacial mixing on thermal conductance. For DC crystalline interfacial regions, we show that the enhancement of conductance by mixing depends on phonon polarization. For instance, mixing increases transmission between TA-TA modes but not between LA-LA modes.

We explain our results within the Landauer theory, where thermal conductance is directly related to the product of the number of conducting channels or modes ($M$) times their average transmission ($T$). We find that 1) the crystal structure determines the relative magnitude of the {\it minimum} of the contacts' modes $M_{min}$ vs. the {\it conserving} modes $M_c$ that conserve the component of phonon momentum transverse to the interface. On the other hand, 2) the interface morphology determines if phonons move through $M_{min}$ for mixed interfaces, or $M_c$ otherwise. Based on these two concepts, we show that the conductance across a mixed interface increases relative to that at a uniform interface when $M_{min}>M_c$, with larger degree of enhancement as the inequality increases. The larger enhancements, seen in SC and in TA branches of DC, are associated with the emergence of voids in the conserving modes ($M_c=0$). Such voids in turn arise when the subbands shift but do not distort with increasing momenta.

We start by deriving an inequality (Eq.~\ref{equinequality}) between the conductance of the mixed and uniform interfaces from the Landauer theory (Sec.~\ref{secprelim}). Then, we describe how the modes (Sec.~\ref{secM}) and transmission (Sec.~\ref{secT}) shape that inequality according to phonon polarization. In Sec.~\ref{secT}, we derive analytical expressions for the transmissions of the scalar SC and FCC systems. For the uniform interface we find a maximum conductance when the junction mass is the arithmetic mean of the contact masses. For the mixed interface, we find that the transmission between phonons that do not conserve transverse wavevector, $k_\perp$, depends on the difference of the contact masses and on the alloy scattering factor, $\alpha(1-\alpha)$ with $\alpha$ the fraction of heavy atoms at the interface.

\section{Landauer Description}\label{secprelim}

Thermal conductance $G^q$ is defined as the ratio between heat flux $q$ and temperature drop $\Delta {\bf T}$.  Within the Landauer theory this quantity can be expressed as \cite{Jeong2012}
\begin{equation}
G^q=\frac{q}{\Delta {\bf T}}=\frac{I^q}{A\Delta {\bf T}}=\frac{1}{A}\int\limits_0^\infty \frac{d\omega}{2\pi}\hbar\omega \frac{\partial N}{\partial {\bf T}} MT,
\label{equGQ}
\end{equation}
where $I^q$ is the heat current, $A$ is the cross-sectional area, $\hbar\omega$ is the energy carried by a phonon, $N$ is the Bose-Einstein distribution, $M$ is the number of propagating modes, which we refer as ``modes'' throughout this paper, and $T$ is the average transmission per mode. For a given contact and frequency $\omega$, the propagating modes are the eigenvectors ($x_n\propto e^{i(kx_n-\omega t)}$) of the equation of motion for the contact with eigenvalue $\omega^2$, with real wavevector $k$ and with group velocity in the transport direction. The product $MT$ equals the sum of the phonon transmissions between modes on the left and right contacts. This quantity can be calculated from non-equilibrium Green's functions (NEGF) as $MT=\text{Trace}\{\Gamma_lG\Gamma_rG^\dagger\}$, with $G$ the retarded Green's function and $\Gamma$ the broadening matrix for the left ($l$) and right ($r$) contacts  \cite{Mingo2003,Datta2005,Wang2008}.

For the uniform interface (Fig.~\ref{fig_cases}c), the symmetry in the transverse direction requires that phonons crossing it conserve their transverse wavevector $k_\perp$. Thus, the nonzero contributions to $MT$ are transmissions $T_{k_{\perp},k_{\perp}}$ between contact modes with the same $k_\perp$. Referring to the number of these transmissions as $M_c$, the conserving modes, and  their average as $T_c$, we can express $MT$ for the uniform interface as
\begin{equation}
MT_{uni}=\sum_{k_{\perp}}T_{k_{\perp},k_{\perp}}=M_cT_c<M_c.
\label{equMTmat}
\end{equation} 
$M_c$ is given by the overlap between the projections of the frequency isosurfaces of the contacts onto the $k_\perp$ plane (Fig~\ref{fig_MT_ToySC}b). Note its role as an upper bound of $MT_{uni}$. Also note that the abrupt interface (Fig.~\ref{fig_cases}a) is a limiting case of the uniform interface.

When we replace the homogeneous interfacial layer of the uniform interface by random contact atoms (Fig.~\ref{fig_cases}b), the atomic disorder breaks the transverse symmetry and allows phonon transmission $T_{k_{\perp},k'_{\perp}}$ between modes that do not conserve $k_\perp$  \cite{Kechrakos1990}. That disorder also decreases the transmission ($\delta T_{c\downarrow}$) between modes that conserve $k_\perp$. We can express MT for the mixed interface as
\begin{align}
MT_{mix}&=\sum_{k_{\perp}}T_{k_{\perp},k_{\perp}} + \sum_{k_{\perp}\neq k'_{\perp}}T_{k_{\perp},k'_{\perp}} \label{equsum} \\ 
&=\underbrace{M_c(T_c-\delta T_{c\downarrow})}_{\text{conserving}}+\underbrace{M_{nc}\delta T_{nc\uparrow}}_{\text{non-conserving}}<M_{\min}, \label{equMTmix}
\end{align}
where $M_{nc}\delta T_{nc\uparrow}$ represents the increase in conductance due to the newly available channels. Note that energy conservation ensures that $MT_{mix}$ is bounded by the minimum of the bulk modes of the contacts: $M_{min}=\min(M_l,M_r)$. Thus we define $M_{nc}=M_{\min}$.

Comparing Eq.~\ref{equMTmat} and \ref{equMTmix}, the conductance of the mixed interface is larger than that of the uniform interface, $G_{mix} > G_{uni}$, if
\begin{equation}
\int_0^\infty \frac{d\omega}{2\pi}\hbar\omega \frac{\partial N}{\partial {\bf T}} M_{min}\delta T_{nc\uparrow}>\int_0^\infty \frac{d\omega}{2\pi}\hbar\omega \frac{\partial N}{\partial {\bf T}} M_c\delta T_{c\downarrow}.
\label{equinequality}
\end{equation}
In other words, $G_{mix} > G_{uni}$ if the gain in conductance by opening new channels that do not conserve $k_\perp$ ($M_{nc}\delta T_{nc\uparrow}$) surpasses the loss in conductance by phonons conserving $k_\perp$ ($M_c\delta T_{c\downarrow}$) over a window set by the cut-off frequency and the temperature.

\section{Minimum vs. Conserving Modes} \label{secM}

We calculate the harmonic conductance of abrupt, uniform and mixed interfaces embedded into four different crystal structures: 1) SC and 2) FCC crystals, where the atomic movements are simplified to a single direction and thus the interatomic force constants (IFCs) are scalars; 3) FCC crystal with IFCs calculated from the Lennard-Jones (LJ) potential; and 4) DC crystal with IFCs calculated from density functional theory (DFT). The interfacial region for each system consists of a monolayer of primitive unit cells (Fig. 1). The same IFCs and lattice constants are used throughout each system to isolate the effect of mass disorder, which has been proven to dominate the scattering of cross-species interactions \cite{Skye2008}. The ratio between the atomic masses of the contacts is 3 for the SC and FCC systems and 2.6 for the DC system, corresponding to the mass ratio of Si and Ge. The conductance is calculated using NEGF, and the details of the simulations and assumptions are given in Appendix~\ref{App_sim_det}. 

\begin{figure}[htb]
	\centering
	\includegraphics[width=70mm]{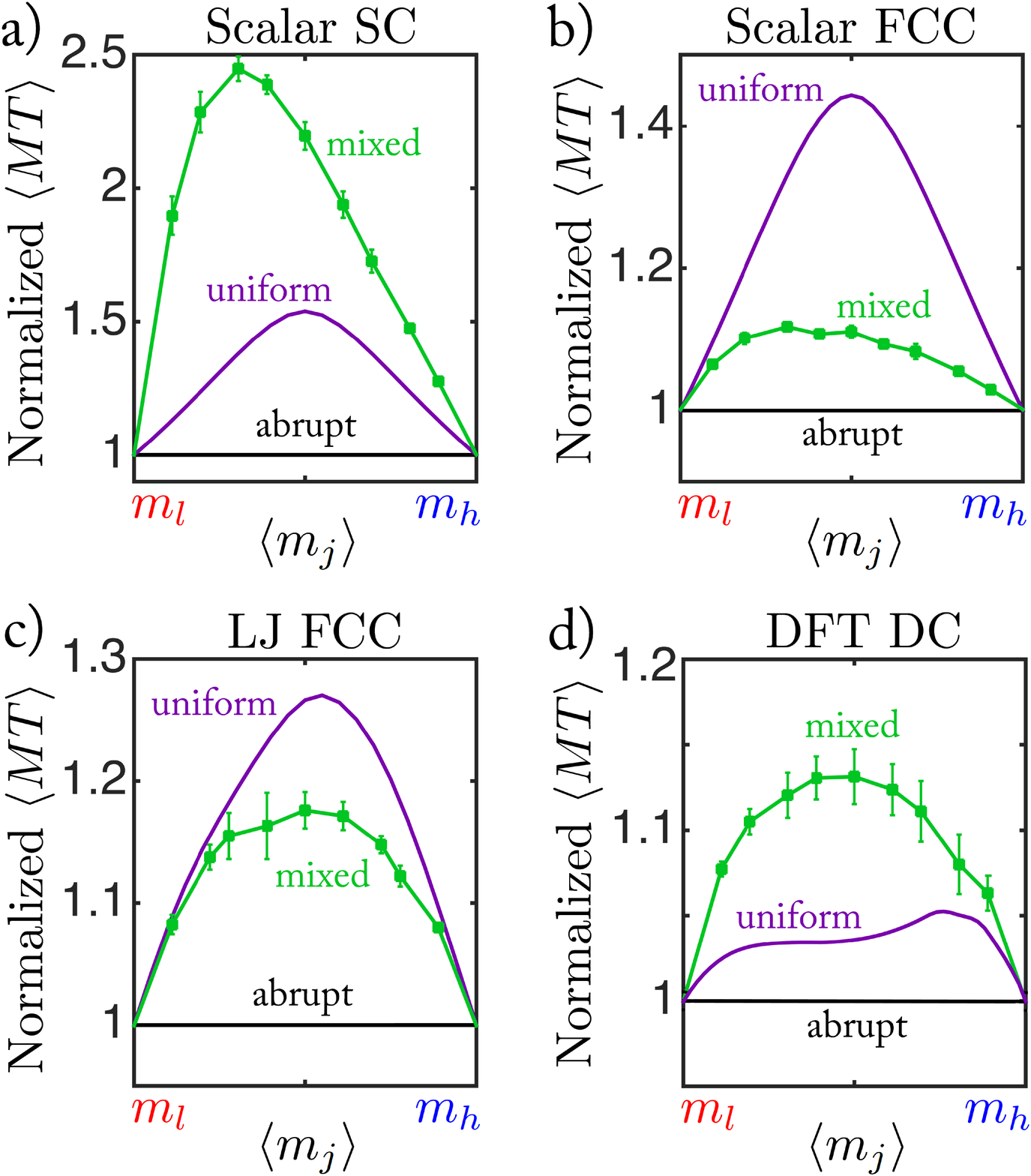}
	\caption{Plot of the average $MT$ normalized by the results for an abrupt interface versus the average mass at the junction layer for the cases described at the beginning of the section. Adding mixing at an abrupt interface enhances the interfacial conductance in all systems simulated in this work, but the extent depends on crystal structure. Compared to uniform interfaces, however, mixing does not always yield an increase in conductance.}
	\label{fig_matvsmix}
\end{figure} 

Figure~\ref{fig_matvsmix} plots the frequency average of $MT$, which one can interpret as conductance without the ``low'' temperature dependence (Eq.~\ref{equGQ}), vs. the average mass at the interfacial layer $\left\langle m_j\right\rangle$ for the different crystal structures. For these systems, adding a unit cell monolayer of mixing to the abrupt interface always enhances the interfacial conductance, but the extent depends on crystal structure. For the scalar SC crystal (Fig.~\ref{fig_matvsmix}a), the maximum relative change of $\langle MT\rangle$ between the mixed and abrupt interfaces is about $((2.5-1)/1\times100\%)=150\%$. However, for the FCC and DC crystals (Fig.~\ref{fig_matvsmix}b, c and d), the relative change of $\langle MT\rangle$ is only about $13\%$. Furthermore, the conductance across a mixed interface does not always increase relative to that at a uniform interface. In fact, it increases for SC and DC crystals but decreases for FCC crystals. 

The increment of conductance from the abrupt interface to the mixed interface (Fig.~\ref{fig_matvsmix}) relies on the atomic extent of the mixing region. For this special case, Eq.~\ref{equinequality} tells us that the gain in conductance by phonons that do not conserve $k_\perp$ surpasses the loss in conductance by phonons conserving $k_\perp$ ($M_{nc}\delta T_{nc\uparrow}>M_c\delta T_{c\downarrow}$). As the extent of the mixing region increases, phonon back scattering increases and transmission decreases. Thus, $\delta T_{nc\uparrow}$ decreases while $\delta T_{c\downarrow}$ increases, making the inequality more difficult to be satisfied. At some point, the inequality stop being true and $G_{mix}$ becomes less than $G_{abr}$, which is the usual experimental outcome \cite{Hopkins2013a}.

\begin{figure}[htb]
	\centering	
	\includegraphics[width=86mm]{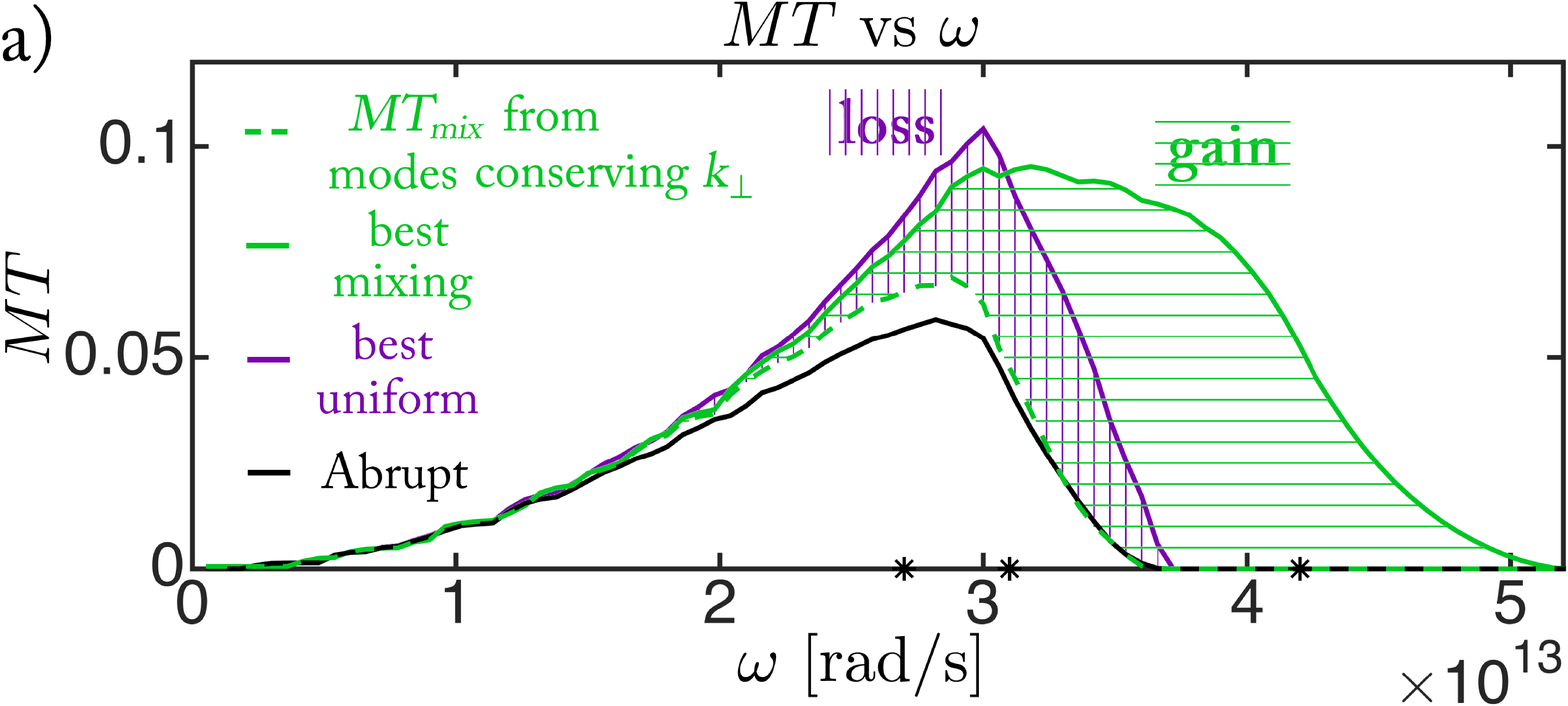}
	\includegraphics[width=86mm]{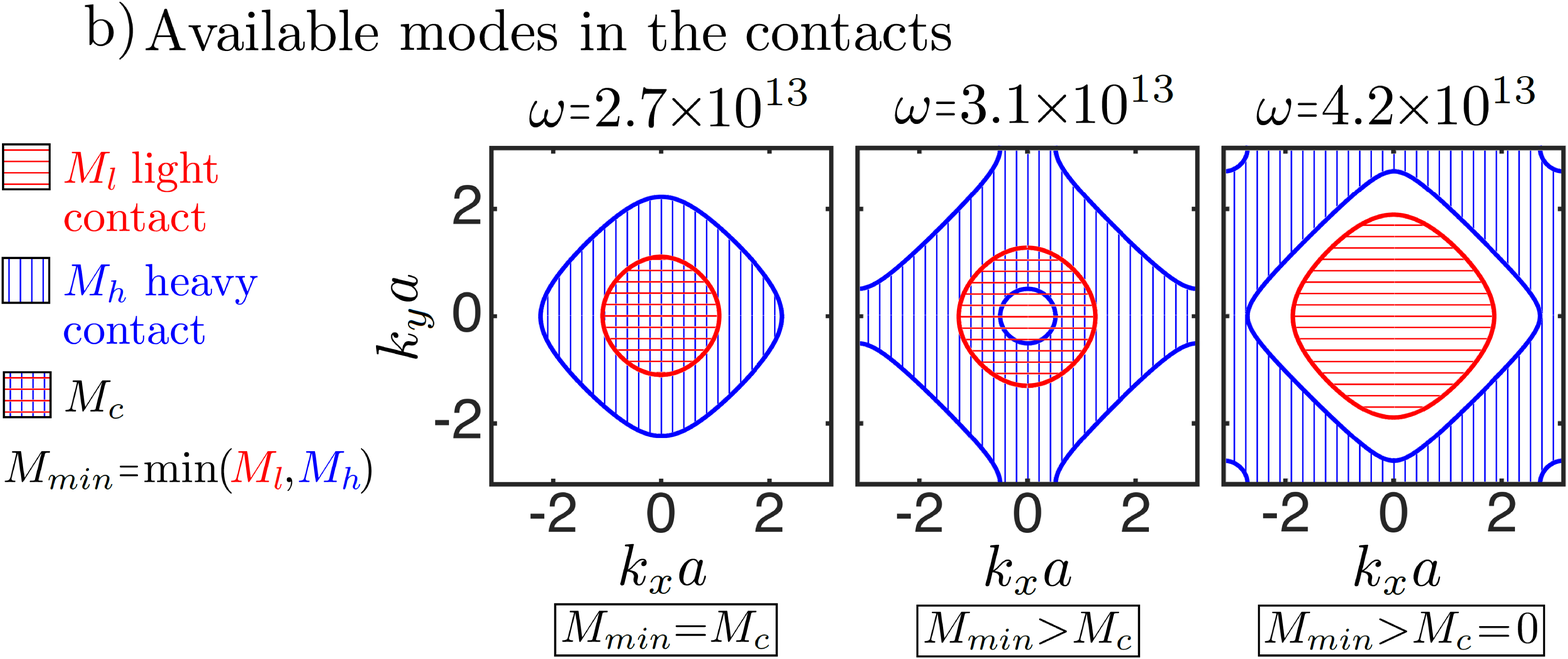}
	\caption{{\bf a)} For SC, the conductance for the mixed interface is larger than the uniform interface because the loss in transmission, $T$, due to the extra scattering brought by the random atoms (area with vertical lines) is dominated by the gain in modes, $M$, coming from transmissions between phonon that do not conserve $k_\perp$ (area with horizontal lines). {\bf b)} While the $MT_{uni}$ spectrum is limited by transmissions between phonon conserving $k_{\perp}$ (overlap region), the extra $MT_{mix}$ spectrum comes from transmissions between phonons that do not conserve $k_{\perp}$.}
	\label{fig_MT_ToySC}
\end{figure} 

For the SC crystal, the large conductance increase of the mixed interface results from the wider $MT$ spectrum (Fig.~\ref{fig_MT_ToySC}a). This extra spectrum comes only from transmissions between modes that do not conserve $k_\perp$. In fact, over that frequency interval, the available contact modes do not overlap (Fig.~\ref{fig_MT_ToySC}b), banning transmissions between modes conserving $k_\perp$. Thus, $M_c=0$ and  $MT_{uni}=0$. Mixing removes the requirement of conserving $k_\perp$, opening $M_{min}$ conduction channels and making $MT_{mix}>0$.

Figure~\ref{fig_MT_ToySC}a shows $MT$ for the mixed, uniform and abrupt interfaces. $MT_{mix}$ is split into the contributions from modes conserving $k_\perp$ and those that do not. This gives us a pictorial representation of Eq.~\ref{equinequality}: $G_{mix}>G_{uni}$ because the $MT$ area gained due to transmissions between modes not conserving $k_\perp$ is larger than the $MT$ area lost due to disorder among the modes that conserve $k_\perp$.

\begin{figure}[htb]
	\centering	
	\includegraphics[width=86mm]{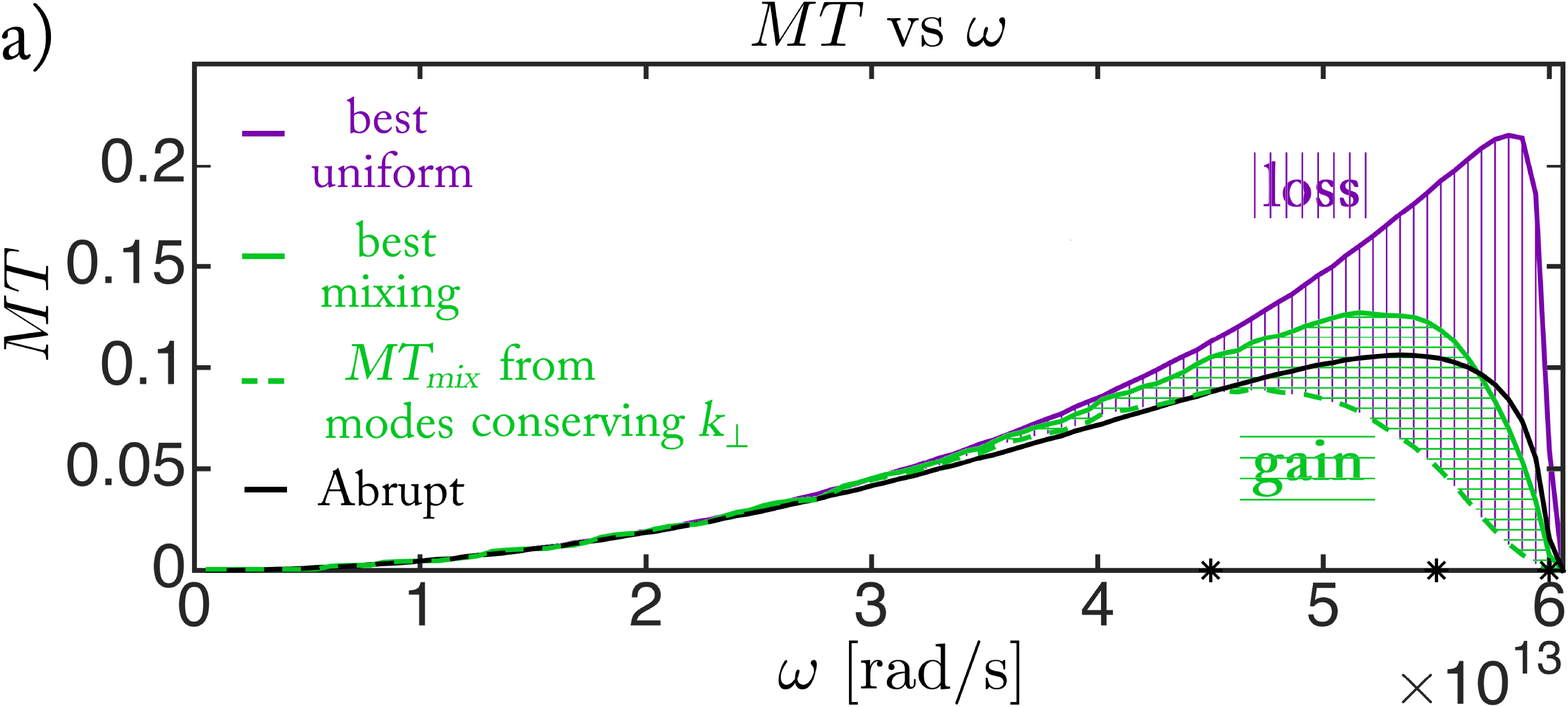}
	\includegraphics[width=86mm]{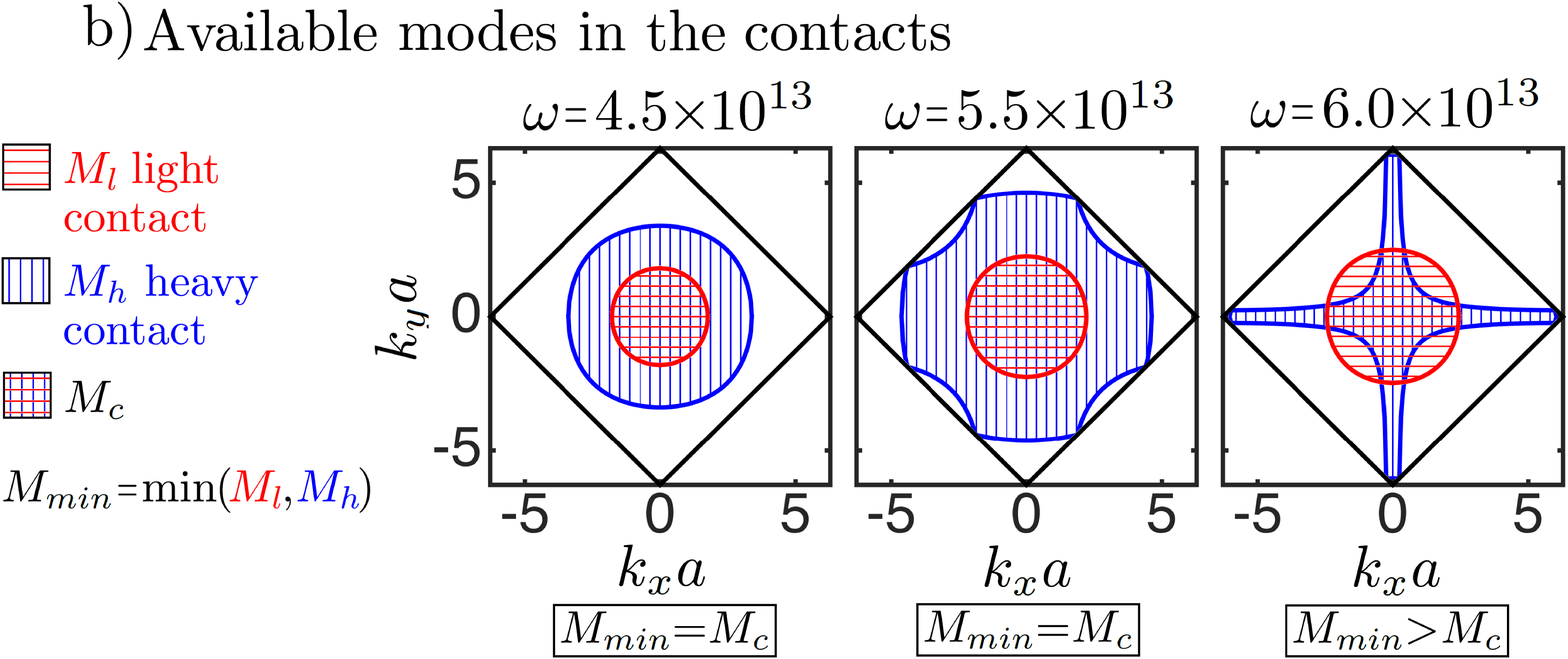}
	\caption{{\bf a)} For FCC, the conductance for the uniform interface is larger than the mixed interface because the loss in $T$ due to the extra scattering brought by the random atoms (area with vertical lines) dominates the gain in $M$ coming from transmissions between phonon that do not conserve perpendicular momentum (area with horizontal lines). {\bf b)} $M_c\approx M_{min}$ over most of the spectrum. Thus, mixing provides little advantage by allowing transmission between modes not conserving $k_\perp$.}
	\label{fig_MT_ToyFCC}
\end{figure} 

A similar pictorial representation for the scalar FCC interfaces is shown in Fig.~\ref{fig_MT_ToyFCC}a. In this case, $G_{uni}>G_{mix}$ because the $MT$ area gained due to transmissions between modes not conserving $k_\perp$ is less than the $MT$ area lost due to disorder among the modes that conserve $k_\perp$. Note that $MT_{uni}$ and $MT_{mix}$ cover the same frequency range, and the overlap of the contacts' modes $M_c$ equals their minimum $M_{min}$ over most of the spectrum (Fig.~\ref{fig_MT_ToyFCC}b). Thus, the accessible modes on the mixed interface $M_{min}$ do not bring any advantage over the existing modes $M_c$ on the uniform interface (Eq.~\ref{equinequality}). The dominant conductance is then decided by the transmission, which in this case favors the loss in the conserving modes over the gain in the non-conserving ones.

From Fig.~\ref{fig_MT_ToySC} and \ref{fig_MT_ToyFCC}, we note that the relative magnitude between $M_{min}$ and $M_c$ plays an important role determining the larger $MT$ between the mixed and uniform interfaces (Eq.~\ref{equinequality}). This is not surprising because of their roles as $MT$ upper bounds for the mixed and uniform cases respectively. We can distinguish three cases: when 1) $M_c\approx M_{min}$, the modes conserving $k_\perp$ reach the physical limit of modes that can carry heat in one of the contacts. Equation~\ref{equinequality} tells us that the transmission alone decides the dominant $MT$, which can be either the uniform or mixed $MT$. For the scalar SC and FCC structures, whenever $M_c\approx M_{min}$ we see that $MT_{uni}>MT_{mix}$ (Fig.~\ref{fig_MT_ToySC} and \ref{fig_MT_ToyFCC}). Therefore the loss in transmission on the conserving modes surpasses the gain in transmission on the non-conserving modes ($\delta T_{c\downarrow}>\delta T_{nc\uparrow}$ in Eq.~\ref{equinequality}). When 2) $M_{min}>M_c$, the dominant $MT$ results from a balance between the added modes that do not conserve $k_\perp$ and the loss in transmission on the modes that conserve $k_\perp$ (Eq.~\ref{equinequality}). For instance, in the SC structure, $MT_{mix}$ becomes larger than $MT_{uni}$ as the ratio $M_{min}/M_c$ increases (Fig.~\ref{fig_MT_ToySC}). When 3) $M_{min}>M_c=0$, $MT_{mix}>MT_{uni}=0$ and  $MT_{mix}$ is only due to transmissions between modes that do not conserve $k_\perp$ as shown by Fig.~\ref{fig_MT_ToySC}. These three criteria may help in the search for interfacial materials where a particular outcome is expected from atomic mixing over the harmonic regime.

For the diamond crystal, the polarization of the incident and transmitted phonons plays an important role in deciding the outcome of the dominant conductance. We give a brief description of the calculation process in Appendix~\ref{apppol}. Figure~\ref{fig_MT_Diamond}a shows that $G_{mix}>G_{uni}$ mostly because of TA phonons in the light contact transitioning to TA and LA phonons in the heavy material. The transmissions between phonons that do not change polarization behave similarly to the scalar crystals. For TA-TA and LA-LA transmissions, $MT_{mix}>MT_{uni}$ when $M_{min}>M_{c}$ and $MT_{uni}>MT_{mix}$ when $M_{min}\approx M_{c}$ (Fig.~\ref{fig_MT_Diamond}b). The transmission between other polarizations will be analyzed in the next section.

\begin{figure}[htb]
	\centering	
	\includegraphics[width=86mm]{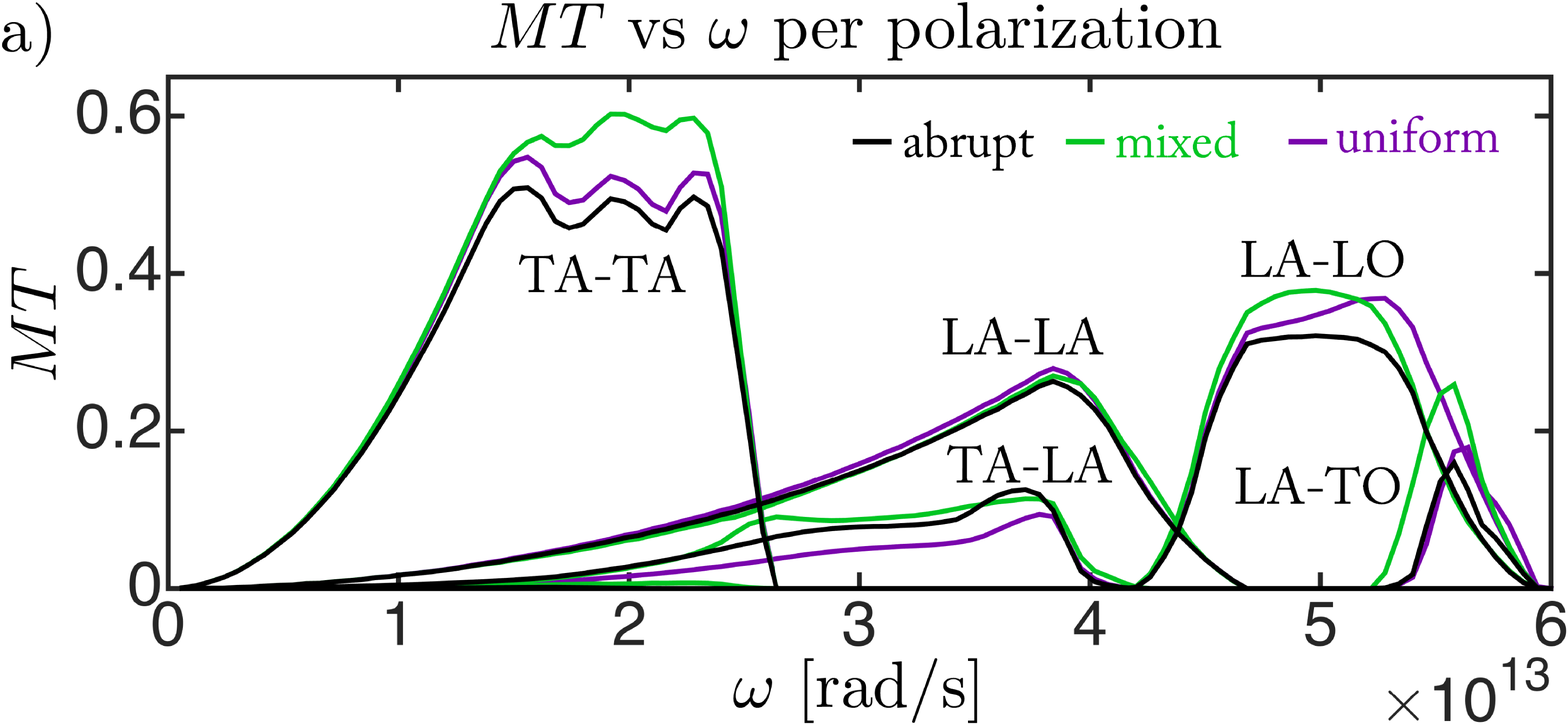}
	\includegraphics[width=86mm]{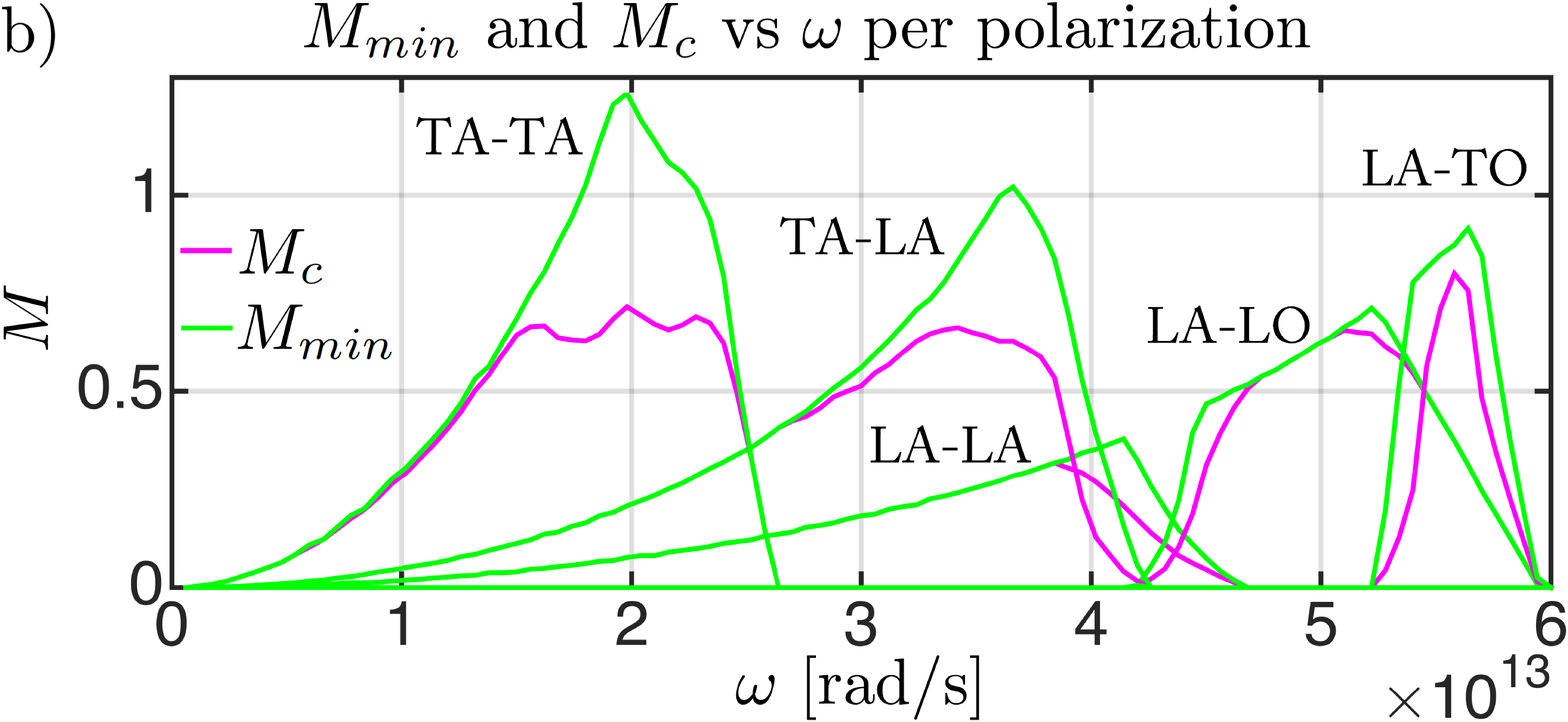}
	\caption{{\bf a)} MT for DC crystal split in polarizations. The conductance of the mixed interface dominates mostly because mixing favors transmissions between TA-TA and TA-LA modes, where the ratio of $M_{min}/M_c$ is larger. {\bf b)} $M_{min}$ and $M_c$ for the different polarization branches.}
	\label{fig_MT_Diamond}
\end{figure}

\begin{figure}[htb]
	\includegraphics[width=86mm]{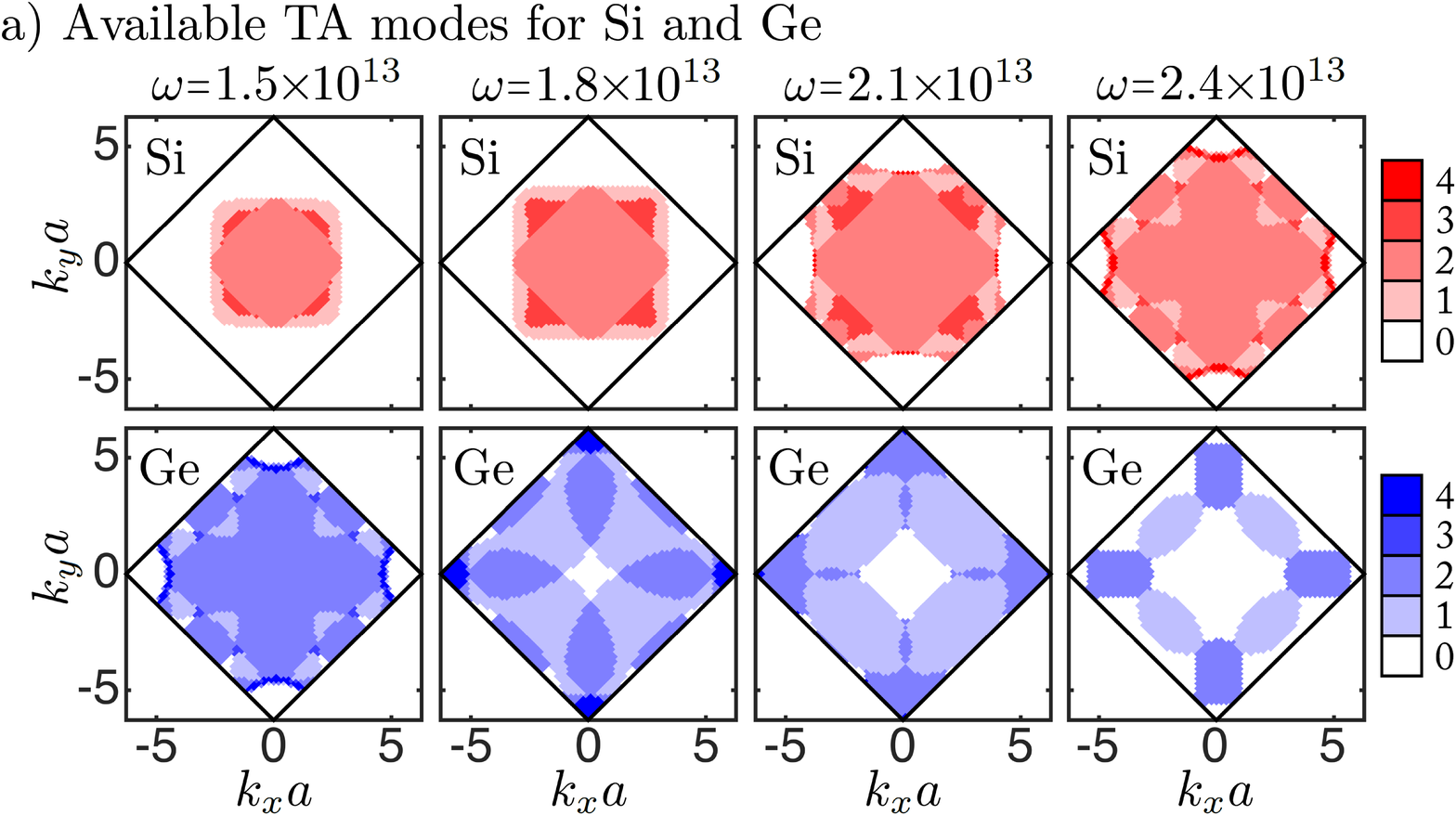}
	\includegraphics[width=86mm]{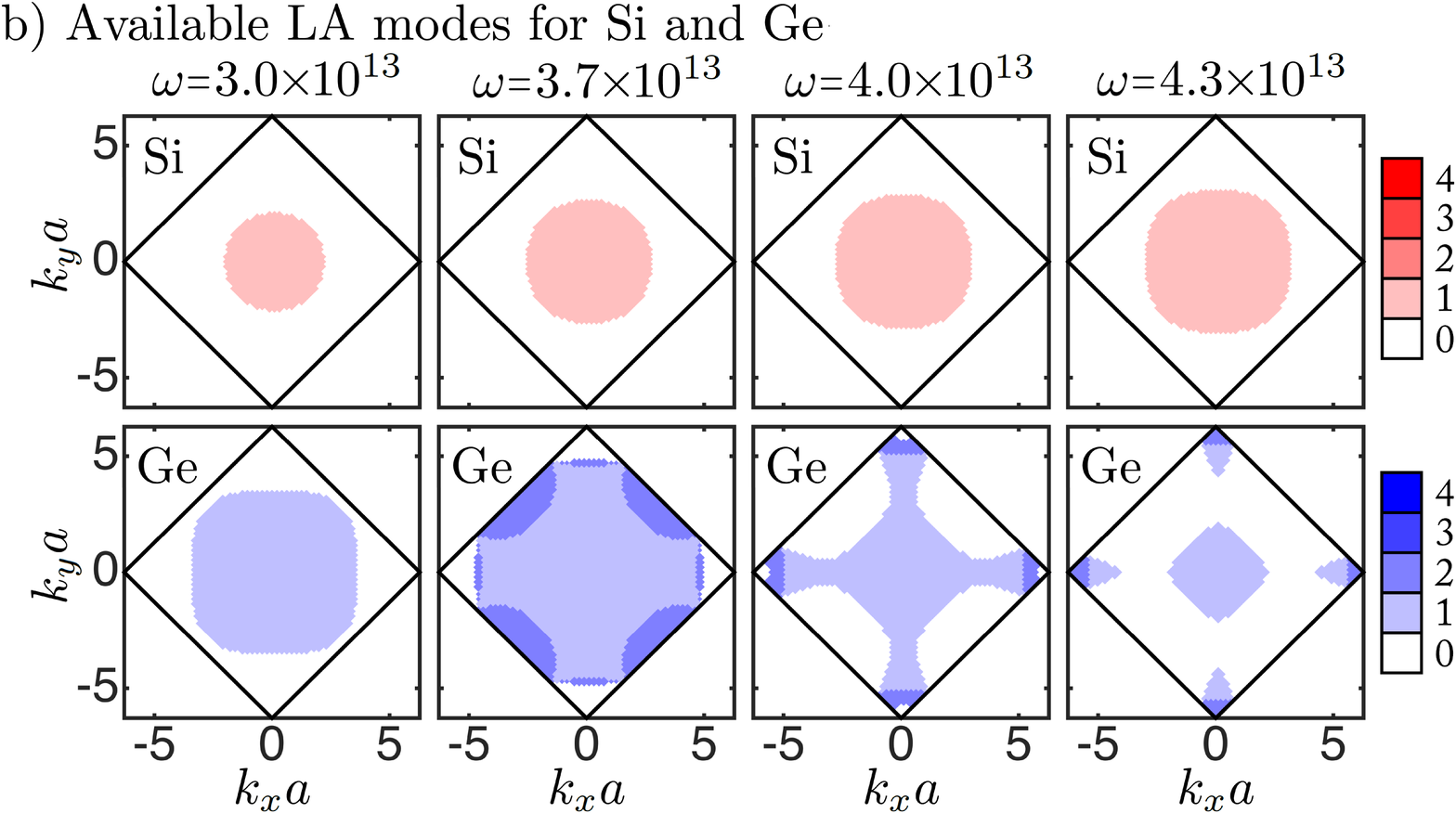}
	\caption{Available modes in the DC contacts for the ({\bf a}) TA and ({\bf b}) LA branches. The modes for SC crystal and the TA branches, where mixing enhances the most the conductance of an abrupt interface, present a central void that enhances the ratio $M_{min}/M_c$.}
	\label{fig_M_projected_Diamond}
\end{figure}

\begin{figure}[htb]
	\includegraphics[width=86mm]{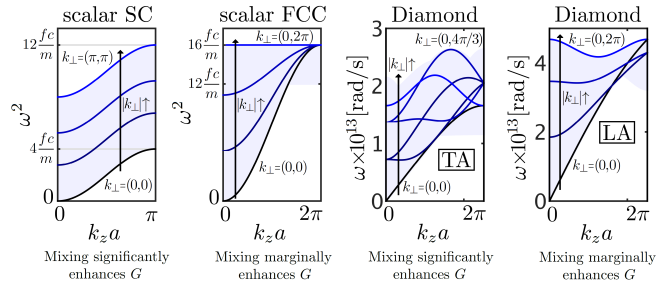}
	\caption{When the shift in the subbands dominate their distortion, the $k_\perp=0$ subband only covers a fraction of the whole band spectrum. This generates a void in the conserving modes, which is seen when mixing significantly enhances the conductance of an abrupt interface. The solid lines represent subbands for different $k_\perp$ and the shaded region is the projected dispersion on $k_za$.}
	\label{figsubands}
\end{figure}

The polarized modes for the diamond crystal uncover an interesting similarity between the modes of the SC and the TA branches in DC and between the modes of the scalar FCC and the LA branch in DC (Fig.~\ref{fig_M_projected_Diamond}). For the cases where mixing significantly enhances conductance (SC and TA-TA branches in DC), we see a common central void in the modes for the heavy contact. This void arises in phonon bands where the $k_\perp\approx 0$ subbands only cover a fraction of the whole band spectrum (Fig.~\ref{figsubands}). Indeed, after the cutoff frequency of those subbands, the $k_\perp\approx0$ modes, or central modes, start to become unavailable. From another point of view, the void originates when the upward shift of the $k_\perp\approx0$ subbands dominate their shrinking as $|k_\perp|$ increases. We see this happening for SC and TA-TA but not for FCC and LA-LA (Fig.~\ref{figsubands}).

For the scalar SC and FCC crystals, the existence of the void can be associated with the independence of interlayer coupling as $k_\perp$ increases. For a SC crystal with atomic mass $m$ and interatomic force constant $f$, the subbands are given by 
\begin{equation}
\omega^2m=f_{on}-2f_{off}\cos(k_za),
\end{equation}
with the onsite coupling $f_{on}=6f-2f\cos(k_xa)-2f\cos(k_ya)$ representing the atomic interactions within a transverse layer of atoms, and the offsite coupling $f_{off}=f$ representing the interaction between layers. As the magnitude of $k_\perp$ increases $f_{on}$ increases, shifting upwards the subband but $f_{off}$ remains constant keeping their width stable. On the other hand, for a FCC crystal the subbands are given by
\begin{equation}
\omega^2m=f_{on} -2f_{off}\cos\left(k_z\frac{a}{2}\right),
\end{equation}
with $f_{on}=12f -4f[\cos(k_xa/2)\cos(k_ya/2)]$ and
$f_{off}=2f[\cos(k_xa/2)+\cos(k_ya/2)]$. As the magnitude of $k_\perp$ increases, $f_{on}$ increases shifting the subbands upward, but at the same time, $f_{off}$ decreases shrinking their bandwidth. 

Although we neglect anharmonicity in this paper, we hypothesize that our main conclusions should hold even when anharmonicity is present. Phonon-phonon interaction enables inelastic transmission of phonons at the interface. However, the transverse symmetry selection rules for $k_{\perp}$  continue to hold. Therefore, phonons crossing an abrupt or uniform interface have to conserve $k_{\perp}$ and are restricted to inelastic jumps within the conserving modes $M_c$. On the other hand, phonons crossing a mixed interface can jump in frequency within the minimum of the contacts' modes $M_{min}$. Thus, we expect a similar relation between the crystal structure, which determines the ratio $M_{min} / M_c$, and the relative magnitude of the conductance for the abrupt, uniform and mixed interfaces. Further studies are required to evaluate the validity of this hypothesis. We also expect a conductance increase for all the systems considered in this work, since anharmonicity allows transmission of phonons with frequencies beyond the elastic limit. Thus as temperature and thereby anharmonicity increases, interfacial thermal conductance increases \cite{Stevens2007,Landry2009}. For some systems with large Debye temperatures anharmonicity can be neglected. For instance, Si/Ge interfaces present a thermal conductance relatively independent of temperature below $500$ K, which indicates that phonon-phonon interactions are not dominant over that temperature range \cite{Landry2009}.

\section{Gain vs. loss in Transmission} \label{secT}

An essential part of the conductance inequality (Eq.~\ref{equinequality}) is the transmission, which can be characterized from our knowledge of $M$ and $MT$. For example at low frequencies, $M_c\approx M_{min}$ and $MT_{abr}\approx MT_{uni} \approx MT_{mix}$, so the transmissions are similar and they only depend on the acoustic mismatch between contacts. Unfortunately most of the spectrum is outside this low frequency regime. 

For the crystals with scalar IFCs over the mid-frequency range, $MT_{uni}>MT_{mix}>MT_{abr}$ as long as $M_c\approx M_{min}$ (Fig.~\ref{fig_MT_ToySC}a and \ref{fig_MT_ToyFCC}a). Therefore, the transmission loss due to disorder for modes that conserve $k_\perp$ dominates the transmission gain from modes that do not conserve $k_\perp$. As frequency increases, thermal energy is carried by shorter wavelength phonons and disorder back scattering accentuates. At some point, it becomes strong enough to reduce $MT_{mix}$ even below $MT_{abr}$ (Fig.~\ref{fig_MT_ToyFCC}a). 

For the DC crystal, the different polarizations available influence the transmission function. For instance, mixing facilitates transmissions between TA-LA modes (Fig.~\ref{fig_MT_Diamond}). This follows from $MT_{mix}>MT_{uni}$ in regions where $M_{min}\approx M_{c}$, which implies $T_{mix}>T_{uni}$. Note that this is the opposite of what we saw for the scalar crystals. Another interesting example shows shifting of transmission between polarizations. Around $\omega\approx5.2\times10^{13}$ rad s$^{-1}$, $MT_{mix}$ for LA-LO decreases while $MT_{mix}$ for LA-TO increases. At this frequency the ratio $M_{min}/M_c$ for LA-TO is larger than for LA-LO so $MT_{mix}$ shifts towards the more favorable condition while conserving energy. In this paper we just scratch the surface of the importance of understanding polarization for interfacial transport, and further studies are required in the topic.

To gain further insights into the transmission, we focus on the crystals with scalar IFCs. For uniform interfaces, Fig.~\ref{fig_matvsmix} shows a conductance maximum when the junction mass is the arithmetic mean (AM) of the contact masses. This follows from a generalization of the same result in 1D interfaces with a single atomic junction \cite{Saltonstall2013,Polanco2013}. By Fourier transforming the transverse coordinates, our 3D problem decouples into a sum of 1D chains with IFCs that depend on the transverse wavevector. For each $k_\perp$ we assume an incident, reflected and transmitted wave and find their amplitudes by solving the equation of motion for the interfacial atom. The transmission $T_{k_\perp}$ follows from the ratio of transmitted over incident current. In this way, $MT$ for the uniform interface is  \begin{equation}
MT_{uni}=\sum_{k_\perp}T_{k_\perp},
\end{equation}
with
\begin{equation}
T_{k_\perp}=\frac{4\Gamma^l_{k_\perp}\Gamma^h_{k_\perp}}{\omega^4\Delta m^2+\left(\Gamma^l_{k_\perp}+\Gamma^h_{k_\perp}\right)^2}.
\label{equmat}
\end{equation}
$\Gamma_{k_\perp}$ is the broadening matrix in NEGF formalism, which reduces to a scalar function when dealing with a single degree of freedom per atom \cite{Datta2005}. This quantity is related to the escape rate of a phonon into the contacts and is given by $\Gamma_{k_\perp}=2\omega \rho v_{k_\perp}$, with $\rho$ the mass density and $v_{k_\perp}$ the frequency dependent group velocity of the mode or subband defined by $k_\perp$. The superscript in $\Gamma_{k_\perp}$ refers to the light $(l)$ and heavy $(h)$ contacts. $\Delta m=m_j-(m_l+m_h)/2$ measures the deviation of the junction mass $m_j$ from the AM of the contact masses. Thus, when $m_j$ is the AM, each $T_{k_\perp}$, $MT_{uni}$ and $G^q_{uni}$ are maxima. Note that $G^q_{uni}>G^q_{abr}$ as long as $m_l<m_j<m_h$, since the abrupt interface is recovered when $m_j=m_h$. Also note that at low frequency, $\omega^4\Delta m^2<<\Gamma_l,\Gamma_h$, the transmissions only depend on the acoustic mismatch between contacts.

A similar generalization from its 1D counterpart \cite{Saltonstall2013,Polanco2013} leads us to conclude that in an abrupt interface where interfacial bonding is the only variable, conductance is maximized when the force constant is the harmonic mean of the contact force constants. In the same fashion, we can generalize other 1D results to 3D interfaces \cite{Polanco2014}.

The conductance maximum derived from Eq.~\ref{equmat} is not valid for tensorial IFCs (Fig.~\ref{fig_matvsmix}). In that case, the amplitudes of the incident and transmitted waves are related through a matrix equation (Eq.~\ref{equC}). $\Delta m \neq 0$ affects both the denominator and the numerator of the transmission, and therefore there is no clear trend when decreasing $\Delta m$. For instance, $\Delta m\neq 0$ might abate the transmission for some polarizations but enhance the transmission between others.

For mixed interfaces, we can approximate $MT_{mix}$ starting from  Eq.~\ref{equMTmixa} (Appendix \ref{Appmixapprox}), the relation between incident and transmitted wave amplitudes at the interface. The heart of the approximation lies on finding the inverse of the matrix $(\tilde{\Delta} +\tilde{Z}_B-\tilde{ Z}_C)^{-1}$, which is a diagonal matrix with tiny off-diagonal elements. These small elements come from Fourier transforming the random mass distribution at the interface. We assume that all these elements are constant, since a random mass distribution contains components in the entire frequency spectrum. Then we estimate their value relating the known real power spectrum with the $k$ space spectrum through Parseval's theorem. Finally, we find the desired inverse using a first order Taylor expansion ($(A+B)^{-1}\approx A^{-1}-A^{-1}BA^{-1}$). With this information, the sum of the transmissions becomes 
\begin{equation}
MT_{mix}=\sum_{k_{\perp}}T_{k_{\perp},k_{\perp}} + \sum_{k_{\perp}\neq k'_{\perp}}T_{k_{\perp},k'_{\perp}},
\label{equMTmix1}
\end{equation}
\begin{equation}
T_{k_{\perp},k_{\perp}}=\frac{4\Gamma^l_{k_\perp}\Gamma^h_{k_\perp}}{\omega^4\left\langle\Delta m\right\rangle^2+\left(\Gamma^l_{k_\perp}+\Gamma^h_{k_\perp}\right)^2}
\label{equMTmix2}
\end{equation} 
and
\begin{equation}
T_{k_{\perp},k'_{\perp}}=\frac{\omega^4(1-\alpha)\alpha(m_l-m_r)^2}{N\Gamma^l_{k_\perp}\Gamma^h_{k_\perp}}T_{k_{\perp},k_{\perp}}T_{k'_{\perp},k'_{\perp}}.
\label{equMTmix3}
\end{equation}
$\left\langle\Delta m\right\rangle$ is the average over the junction masses, $N$ is the number of atoms in the cross section and $\alpha$ is the fraction of heavy atoms at the interface. Equation~\ref{equMTmix3} suggests that the transmission between modes that do not conserve $k_\perp$, $T_{k_{\perp},k'_{\perp}}$, is proportional to the square of the difference between the atomic masses of the contacts, $(m_l-m_r)^2$, to the alloy scattering factor, $(1-\alpha)\alpha$, and to some function of the acoustic properties of the contacts. The equation does not capture the decrease in transmission among the modes that conserve $k_{\perp}$ due to disorder. It also over predicts the contribution from transmissions that do not conserve $k_\perp$ and does not capture their asymmetric bias as a function of junction mass (Fig.~\ref{fig_matvsmix}). In spite of that, it provides a sense for the expected conductance enhancement by mixing and insight on how to build the transmission between different modes, which is an important step forward towards qualitative understanding of interfacial conductance.

\section{Conclusion}

In this manuscript we quantify the role of crystal structure and interface morphology on the interface thermal conductance. We show that the crystal structure (SC: simple cubic, FCC: face centered cubic, or DC: diamond cubic) determines the relative magnitude of the {\it minimum} of the contacts' modes $M_{min}$ vs. the {\it conserving} modes $M_c$ that conserve the component of phonon wavevector transverse to the interface $k_{\perp}$. On the other hand, the interfacial morphology (abrupt, uniform: with an added homogeneous layer, or mixed: with atomic disorder) determines if phonons can move through $M_c$ or $M_{min}$. 

We find that adding  a unit cell monolayer of mixing to an abrupt interface enhances the interfacial conductance, but the degree depends on the ratio $M_{min}/M_c$. In particular for a scalar FCC crystal, where $M_{min}\approx M_c$, the conductance of a mixed interface increases relative to that of an abrupt interface by 13\%. This modest enhancement comes from a balance between the new accessible modes and the extra scattering created by disorder. For a SC crystal, the relative conductance increment from the abrupt interface to the mixed interface is $\sim ~150\%$. This large enhancement comes from a region where there are available modes but they do not overlap ($M_{min} > M_c = 0$) because of a central void of modes in the Brillouin zone. The void appears when the upward shift in the subbands dominate their shrinking as $k_\perp$ increases, which happens due to the independence of interlayer coupling from $k_\perp$. 

For a DC crystal, we find that the effect of mixing depends on the polarization. In particular, mixing increases transmissions between TA branches but not between LA branches. The modes for the TA branches present a central void similar to what we saw in the SC crystal. This suggests that materials with modes containing central voids are prone to high conductance enhancement by mixing. We also find that the conductance across a mixed interface does not always increase relative to that at a uniform interface. In fact, it increases for SC and DC but decreases for FCC, and is once again correlated with the ratio $M_{min}/M_c$. This suggests that the commonly invoked virtual crystal approximation alternatively overestimates or underestimates the effect of interfacial mixing on thermal conductance.

\begin{acknowledgments}
C.A.P., J.Z. and A.W.G. are grateful for the support from NSF-CAREER (QMHP 1028883) and from NSF-IDR (CBET 1134311). They also acknowledge XSEDE (DMR130123) for computational resources, and the School of Engineering and Applied Sciences at University of Virginia that covered our registration fees for a Quantum Espresso workshop. R. R., N. Q. L., and P. M. N. gratefully acknowledge support from the AFOSR (FA9550-14-1-0395). P. E. H. appreciates support from the Office of Naval Research (N00014-13-4-0528).
\end{acknowledgments}

\appendix

\section{Simulation details}\label{App_sim_det}

Each interface consists of two contacts joined by a layer of primitive unit cells (Fig.~\ref{fig_cases}). We find the interfacial thermal conductance using NEGF and assume that the crystal structure, lattice constant $a$ and IFCs are invariant throughout each system. This commonly used simplification \cite{Schelling2002,McGaughey2006,Stevens2007,English2012} provides an easy way to study thermal conductance through vibrationally mismatched interfaces. Moreover, the simplification is well suited for Si/Ge interfaces because the IFCs of these materials are very similar \cite{Tian2012} and therefore the difference in atomic mass is a dominant scattering mechanism \cite{Skye2008}. The ratio between the atomic masses of the contacts is 3 ($m_l=40$ amu and $m_h=120$ amu) for all the systems but the diamond crystal, where we use the masses of Si and Ge. 

The IFCs for the scalar SC and FCC interfaces are built considering only nearest neighbor interactions described by a force constant of $45$ N / m. Assuming $a=5$ \AA, the thermal conductance for the abrupt interface is given by $G_{abr}=7.5$ MW m$^{-2}$ K$^{-1}$  for SC and $G_{abr}=44.3$ MW m$^{-2}$ K$^{-1}$ for FCC at a temperature of $300$ K. Note that the value for FCC is $\sim6$ times larger than for SC because the FCC crystal has twice the number of atoms per cross sectional area and its MT is $\sim3$ times larger (Fig.~\ref{fig_MT_ToySC}a and \ref{fig_MT_ToyFCC}a). 

For the FCC LJ interfaces, the IFCs are extracted from the Lennard-Jones potential using $\epsilon=0.0503$ eV, $\sigma=3.37$ \AA\ and a cut-off distance of $2.5\sigma$. This potential generates interactions up to fifth-nearest atomic neighbors and corresponds to an equilibrium lattice constant of $a=5.22$ \AA. The  potential is chosen to be identical to that used by English et al. \cite{English2012} to have a point of reference for benchmarking. In fact, we checked the consistency of our IFCs by comparing the phonon dispersions and densities of states against the reference. The conductance for the abrupt interface is $G_{abr}=57.8$ MW m$^{-2}$ K$^{-1}$ at a temperature of $147$ K. Our non-equilibrium molecular dynamics (NEMD) simulations predict a larger $G_{abr}=97.41$ MW m$^{-2}$ K$^{-1}$ at a temperature of $30$ K due to anharmonic transmission of phonons beyond the cut off frequency of the heavy material. Note that very low temperature NEMD results, which are classical and mostly harmonic, should tend to high temperature NEGF results, where the Bose-Einstein distribution approaches the classical limit.

For the DC crystal we use the IFCs from silicon extracted using Quantum Espresso, which is a software package for performing calculations using density functional perturbation theory, that has successfully predicted and matched experimental Kapitza conductance and thermal conductivity without any fitting parameters \cite{Giannozzi09}. In this calculation, we used local density approximation (LDA) of Perdew and Zunger \cite{Perdwe1981} with direct fitting. The cutoff energy for the planewave kinetic energy is 30 Ryd, while the k sampling is $4\times4\times4$ with Monkhorst-Pack method. We also considered $4\times4\times4$ q points when calculating the dynamical matrix. The lattice constant for silicon is found to be 5.398 \AA. Our parameters were chosen after carefully satisfying convergence tests, and the dispersion of silicon matches the experimental data quite well. For simplicity in the calculations, we only consider interactions up to the second nearest neighbor. Our simulations predict $G_{abr}=242.5$ MW m$^{-2}$ K$^{-1}$ at $300$ K. To check our code, we simulate the same interface with IFCs extracted from Stillinger-Weber potential and obtain $G_{abr}=276.6$ MW m$^{-2}$ K$^{-1}$ at $300$ K, which is comparable to the $G_{abr}=280$ MW m$^{-2}$ K$^{-1}$ reported by Tian et. al. \cite{Tian2012} for the same interface, potential and temperature. Those values are within 15\% of the ones obtained using lattice dynamics and NEMD calculations $G_{abr}\approx310$ MW m$^{-2}$ K$^{-1}$ at $300$ K \cite{Zhao2005,Landry2009}.

All our $MT$ calculations are done in transverse wavevector space ($k_\perp$-space) to simplify the 3D problem into a sum of 1D independent problems. For the abrupt and uniform interfaces each 1D chain consists of primitive unit cells. For the mixed interfaces we increase the size of the unit cell and randomly choose the atoms at the junction layer according to the desired average mass. The unit cell for SC has 36 atoms, for FCC has 32 atoms, for FCC LJ has 18 atoms and for diamond has 36 atoms. For the diamond crystal we also simulated 16 and 64 atoms and did not see appreciable changes in the results. Based on this, the results for FCC LJ might change less than 5\% if we increase the number of atoms. The $MT$ for scalar SC and FCC agree with the $MT$ obtained using Eq.~\ref{equMTmixa}. For each mixed interface, we report the average over more than 12 independent calculations and in Fig.~\ref{fig_matvsmix} we also report the standard deviation.

To split the contribution of $MT_{mix}$ from the modes that conserve and do not conserve $k_\perp$ (dashed line in Fig.~\ref{fig_MT_ToySC}a and \ref{fig_MT_ToyFCC}a), we find the transmission directly from Eq.~\ref{equMTmixa} in a system with $40 \times 40$ atoms in the cross section and periodic boundary conditions.  Our results show the average over 12 independent simulations of random distributions of atoms at the junction.

To calculate propagating modes for a contact we simulate an ``interface'' where the leads and junction are the same material. In this case $T=1$ because there is no interface and $MT=M$. The dispersions in the scalar SC and FCC crystals are simple enough that we found the propagating modes analytically by projecting the frequency iso-surface onto the $k_x,k_y$ plane.

\section{Polarization-Resolved Transmission}\label{apppol}

To find the transmission resolved by polarizations we start by 1) choosing a frequency $\omega$ for which we identify all the propagating and evanescent modes of both contacts. This is done by solving a generalized eigenvalue problem as explained by Wang et. al. (Sec. 2.2.2 of \cite{Wang2008}). Then we 2) assign a polarization to each of the propagating modes. That is, we find the dispersion branch to which each mode belongs. This is done by moving in small wavevector increments from $k_0$, a fixed wavevector where we know the correspondence between eigenvalues (frequencies), eigenvectors (polarizations) and branches, to $k'$, the wavevector of the phonon we want to label. In each step we calculate the eigenvectors of adjacent $k$ grid points and project ones into the others. Then according to the maximum projection between eigenvectors we assign a branch to  each of the eigenvalues and eigenvectors of the next grid point. Once we assigned a label to each propagating mode we 3) find the response around the interface to an incident mode from the left contact. This is done using the Green's function of the system, which is the impulse response of the system, by exciting the system with a superposition of impulses that resemble the mode. Then we 4) project the part of the response at the right contact onto the modes of that contact. At this point we have the amplitude of the impinging and transmitted modes. Finally, we 5) find the current carried by each mode and the transmission between modes, which we label according to the labels of the modes involved. For mixed interfaces we have to unfold the branches of the supercell to be able to identify the polarization and label consistently with primitive unit cell polarizations.

\section{Transmission for Mixed Interface}

Our aim is to solve the scattering problem of a wave impinging on an interface to obtain Eq.~\ref{equMTmix1}, \ref{equMTmix2} and \ref{equMTmix3}. To this end we 1) assume incident, reflected and transmitted waves and find an equation relating their amplitudes. Then we 2) approximate that equation to find an analytical solution. Finally we 3) find the transmission from the ratio between transmitted and incident currents and sum them up to get $MT$. 

\subsection{Equation Relating Amplitudes}

\begin{figure}[t]
	\centering
	\includegraphics[width=86mm]{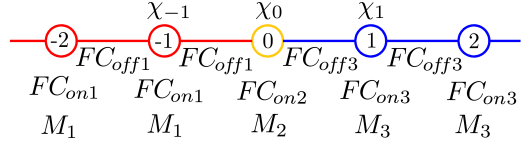}
	\caption{System split into sites in the transport direction.}
	\label{fig1Dchain}
\end{figure}

Consider a system split into sites in the transport direction (Fig.~\ref{fig1Dchain}) and described by the equation of motion
\begin{equation}
\omega^2M_{sys}\chi=F_{sys}\chi,
\end{equation}
with $M_{sys}$ the mass matrix of the system
\begin{equation}
M_{sys}=\begin{pmatrix}
\ddots & & & & \\
& M_1 & & & \\
& & M_2 & & \\
& & & M_3 & \\
& & & & \ddots
\end{pmatrix},
\end{equation}
$F_{sys}$ the force constant matrix of the system
\begin{equation}
F_{sys}=\begin{pmatrix}
\ddots & & & & \\
& F_{on1} & F_{off1} & & \\
& F_{off1}^\dagger & F_{on2} & F_{off3} & \\
& & F_{off3}^\dagger & F_{on3} & \\
& & & & \ddots
\end{pmatrix}
\end{equation}
and $\chi$ the vector containing the displacements from equilibrium of each of the atoms of the system. The equation of motion for the interfacial site is given by 
\begin{equation}
\omega^2M_2\chi_0=F_{on2}\chi_0+F_{off1}^\dagger\chi_{-1}+F_{off3}\chi_1
\label{equmotint}
\end{equation} 
Because of the periodicity of the contacts, plane waves of the form $\chi_n=X_{j}e^{ik_{j}na}$ (Bloch states) satisfy the equation of motion for any contact site if $X_j$, the polarization vector, satisfies  
\begin{equation}
\omega^2MX_j=\left[F_{on}+F_{off}^\dagger e^{-ik_ja}+F_{off}e^{ik_ja}\right]X_j.
\label{equmc}
\end{equation} 
In terms of these plane waves we assume a solution for the system of the form
\begin{equation}
\chi_n=\sum_{k_1^+}A_{k_1^+}X_{k_1^+}e^{ik_{1}^+na}+\sum_{k_1^-}B_{k_1^-}X_{k_1^-}e^{-ik_1^-na}
\label{equasl}
\end{equation}  
for $n\leq0$, and for $n\geq0$
\begin{equation}
\chi_n=\sum_{k_3^+}C_{k_3^+}Y_{k_3^+}e^{ik_3^+na},
\label{equasr}
\end{equation} 
where $+$ and $-$ refer to plane waves propagating to the right or left, $X$ and $Y$ refer to the polarizations on the left and right contacts. We replace the assumed solution (Eq.~\ref{equasl} and \ref{equasr}) into the equation of motion at the interface (Eq.~\ref{equmotint}). For the factor $F_{off1}^\dagger\chi_{-1}$, we split each $F_{off1}^\dagger e^{ik_1^\pm a}$ into Hermitian and anti-Hermitian parts. We replace the Hermitian part in favor of $\omega^2M_1-F_{on1}$ using  Eq.~\ref{equmc} and reorganize the anti-Hermitian part in matrix notation to get
\begin{equation}
F_{off1}^\dagger\chi_{-1}=\frac{\omega^2M_1-F_{on1}}{2}\chi_0-Z_AV_{1+}A-Z_BV_{1-}B,
\label{equp1}
\end{equation}
with $V_{1+}$ and $V_{1-}$ the matrices whose columns are the polarizations $X_{k_1^+}$ and $X_{k_1^-}$ respectively and with 
\begin{equation}
Z_A=\frac{F_{off1}V_{1+}\lambda_{1+}V_{1+}^{-1}-F_{off1}^\dagger V_{1+}\lambda_{1+}^{-1}V_{1+}^{-1}}{2},
\label{equZA}
\end{equation}
\begin{equation}
Z_B=\frac{F_{off1}V_{1-}\lambda_{1-}V_{1-}^{-1}-F_{off1}^\dagger V_{1-}\lambda_{1-}^{-1}V_{1-}^{-1}}{2},
\end{equation}
\begin{equation}
\lambda_{1\pm}=\begin{pmatrix}
e^{ik_{11}^\pm a} & & & \\
& e^{ik_{12}^\pm a} & & \\
& & e^{ik_{13}^\pm a} & \\
& & & \ddots
\end{pmatrix},
\end{equation}
where the second subindex of $k_{11}^\pm$ run over the possible $k_{1}^\pm$. In a similar way we get that
\begin{equation}
F_{off1}\chi_{1}=\frac{\omega^2M_3-F_{on3}}{2}\chi_0+Z_CV_{3+}C
\label{equp2}
\end{equation}
with
\begin{equation}
Z_C=\frac{F_{off3}V_{3+}\lambda_{3+}V_{3+}^{-1}-F_{off3}^\dagger V_{3+}\lambda_{3+}^{-1}V_{3+}^{-1}}{2}.
\end{equation}

Equating Eq.~\ref{equasl} and \ref{equasr} at $n=0$ and putting Eq.~\ref{equp1} and \ref{equp2} into Eq.~\ref{equmotint} we get the following set of equations
\begin{equation}
\chi_0=V_{1+}A+V_{1-}B=V_{3+}C,
\end{equation}
\begin{equation}
\Delta \chi_0=-Z_AV_{1+}A-Z_BV_{1-}B+Z_CV_{3+}C,
\end{equation}
with 
$$\Delta=\omega^2\left[M_2-\frac{M_1+M_3}{2} \right]-\left[F_{on2}-\frac{F_{on1}+F_{on3}}{2}\right]$$
From there we can derive the coefficients for the transmitted waves
\begin{equation}
C=V_{3+}^{-1}\left(\Delta+Z_B-Z_C\right)^{-1}\left(Z_B-Z_A\right)V_{1+}A
\label{equC}
\end{equation}

\subsection{Approximation of Amplitudes Equation}\label{Appmixapprox}

Imagine that each site on Fig.~\ref{fig1Dchain} consists of a cross sectional plane of atoms in the mixed interface (Fig.~\ref{fig_cases}b). For the SC and FCC scalar systems, the force constants are invariant in the transport direction, periodic in the transverse direction and scalar between atoms. Thus, $V=V_{1+}=V_{1-}=V_{3+}$. $V$ is the matrix associated with a Fourier transformation into the transverse $k$-space, whose columns are plane waves defined by $k_\perp$ over the $N$ atomic positions $r_n$ in a cross sectional plane
\begin{equation}
V=\frac{1}{\sqrt{N}}\begin{pmatrix}
| & | &  \\
e^{ik_{\perp1}r_n} & e^{ik_{\perp2}r_n} & \cdots \\
| & | & 
\end{pmatrix}.
\end{equation}
Using this information we simplify the relation between the impinging and transmitted waves (Eq.~\ref{equC}) as
\begin{equation}
C=\left(\tilde{\Delta} +\tilde{Z}_B-\tilde{ Z}_C\right)^{-1}\left(\tilde{Z}_B-\tilde{Z}_A\right)A
\label{equMTmixa}
\end{equation}
where the tilde means the matrix in Fourier space, i.e. $\tilde{Z}_A=V^\dagger Z_AV$. Because of transverse periodicity, all the matrices in Eq.~\ref{equMTmixa} are diagonal except 
\begin{equation}
\left[ \tilde{M}_2\right]_{i,j}=\frac{1}{N}\sum_n [M_2]_{n,n} e^{i(k_{\perp j}-k_{\perp i})\cdot r}
\end{equation}
For $i=j$ the term reduces to the average if the interfacial masses
\begin{equation}
\left[\tilde{M}_2\right]_{i,i}=\left\langle M_2 \right\rangle=(1-\alpha)m_{l}+\alpha m_{h},
\end{equation}
where $\alpha$ is the fraction of heavy atoms at the interfacial layer.
For $i\neq j$ we are calculating a frequency component of a random distribution of masses, which should spam over all the $k_\perp$ spectrum. Thus we assume that all the off diagonal components of $\tilde{M}_2$ have the same magnitude. We estimate the value using  Parseval's theorem, the power spectrum in real space and the transformation of the interfacial mass function at $k_\perp=0$ 
\begin{equation}
\left|\tilde{M}_{i,j}\right|=\sqrt{\frac{(1-\alpha)\alpha}{N-1}}\left|m_l-m_h\right|.
\end{equation}
Plugging this simplification and $Z_B=-Z_A$ into Eq.~\ref{equMTmixa} our problem reduces to solve
\begin{equation}
C=-2\left[\begin{pmatrix}
\zeta_{k_{\perp1}} & \epsilon  & \\
\epsilon & \zeta_{k_{\perp2}}  & \\
& &  \ddots
\end{pmatrix}\right]^{-1}
\begin{pmatrix}
\tilde{Z}_{Ak_{\perp1}} & & \\
& \tilde{Z}_{Ak_{\perp2}} & \\
& & \ddots
\end{pmatrix} A,
\label{equCA}
\end{equation}
with 
$$\zeta_{k_\perp}=\omega^2\left(\left\langle M_2 \right\rangle-\frac{m_{h}+m_{l}}{2}\right)-(\tilde{Z}_{Ak_\perp}+\tilde{Z}_{Ck_\perp})$$
$$\epsilon=\omega^2\sqrt{\frac{(1-\alpha)\alpha}{N-1}}\left|m_l-m_h\right|.$$
$\epsilon$ is small since it is inversely proportional to $\sqrt{N-1}$, so we approximate the inverse of the matrix using the first order of its Taylor expansion $(A+B)^{-1}\approx A^{-1}-A^{-1}BA^{-1}$ with $A$ being the diagonal part and $B$ the rest. Finding the inverse and solving Eq.~\ref{equCA} we get that
\begin{equation}
C=QA
\end{equation}
with
\begin{equation}
Q_{ln}=\begin{cases} 
\frac{-2Z_{Ak_{\perp n}}}{\zeta_{k_{\perp n}}}
 &\mbox{if } l=n \\ 
\frac{\epsilon2Z_{Ak_{\perp n}}}{\zeta_{k_{\perp l}}\zeta_{k_{\perp n}}} & \mbox{if } l\neq n \end{cases} 
\end{equation}
Where $Q_{ln}$ relates the amplitude $A_n$ of the $n$ incident mode with amplitude $C_{l}$ of the $l$ transmitted mode.

\subsection{Find the Transmission}

Now that we know the coefficients we can calculate the transmission from mode $A_n$ to mode $C_l$ by dividing the transmitted by the incident current \cite{Polanco2013, Polanco2014}
$$T_{ln}=\frac{\Gamma_{rk_l}}{\Gamma_{lk_n}}\left|\frac{C_l}{A_n}\right|^2=\frac{\Gamma_{rk_l}}{\Gamma_{lk_n}}\left|M_{ln}\right|^2$$
to obtain
\begin{equation}
T_{ln}=\begin{cases} 
\frac{\Gamma_{lk_{\perp n}}\Gamma_{rk_{\perp n}}}{\left|\zeta_{k_{\perp n}}\right|^2}
 &\mbox{if } l=n \\ 
\frac{\epsilon^2\Gamma_{lk_{\perp n}}\Gamma_{rk_{\perp l}}}{\left|\zeta_{k_{\perp l}}\right|^2\left|\zeta_{k_{\perp n}}\right|^2} & \mbox{if } l\neq n \end{cases} 
\end{equation}
with 
$$\zeta_k=\omega^2\left[\left\langle m_n\right\rangle -\frac{m_l+m_r}{2}\right]+i\left[\frac{\Gamma_{lk_n}}{2}+\frac{\Gamma_{rk_n}}{2}\right].$$
Here we replace 
$$Z_{Ak_\perp}=-i\frac{\Gamma_{lk_\perp}}{2}\ \ \ \ Z_{Ck_\perp}=-i\frac{\Gamma_{rk_\perp}}{2},$$
which is true only for the propagating modes and therefore it works only when both of the modes involved in $T_{ln}$ are propagating, i.e. when $T_{ln}\neq0$.

Then the $MT$ per unit cell is 
\begin{align*}
MT_{puc}&=\frac{1}{N}\sum_n \frac{\Gamma_{lk_n}\Gamma_{rk_n}}{\left|\zeta_{k_n}\right|^2} + \frac{1}{N}\sum_{l\neq n}\frac{\epsilon^2\Gamma_{lk_n}\Gamma_{rk_l}}{\left|\zeta_{k_l}\right|^2\left|\zeta_{k_n}\right|^2}
\end{align*}
and from there Eq.~\ref{equMTmix1}, \ref{equMTmix2} and \ref{equMTmix3} follow.

\bibliography{bibpaper}

\end{document}